\documentclass[useAMS,usenatbib]{mn2e}
\usepackage{times}
\title[The Effect of Toroidal Magnetic Field on Thickness Hot Flow ]{The Effect of Toroidal Magnetic Field on Thickness of  a Viscose-Resistive Hot Accreting Flow}
\author[Samadi M., Abbassi S.  $\&$ Khajavi M.]{
Samadi M. $^{1}$\thanks{$Email: M_-Samadi_-M@yahoo.com$},  Abbassi S.$^{1,2}$ \thanks{$Email: Abbassi@ipm.ir$}, Khajavi M.$^{1} $ \\
$^{1}$Department of Physics, School of Sciences, Ferdowsi University of Mashhad, Mashhad, 91775-1436, Iran\\
$^{2}$School of Astronomy, Institute for Studies in
Theoretical Physics and Mathematics, P.O.Box 19395-5531, Tehran, Iran}
\date{}
\begin{document}
\pagerange{\pageref{firstpage}--\pageref{lastpage}} \pubyear{2012}

\maketitle \label{firstpage}

\begin{abstract}
By taking into account the effect of toroidal magnetic field and its correspond heating, we determine the thickness of
advection-dominated accretion flows. We consider an axisymetric, rotating, steady viscous-resistive, magnetized accretion flow under an advection dominated stage. The dominant mechanism of energy dissipation is assumed to be turbulence viscosity and magnetic diffusivity. We adopt a self-similar assumption in the radial direction to obtain the dynamical quantities, that is, radial, azimuthal, sound and Alfv\' en velocities. Our results show the vertical component of magnetic force acts in the opposite direction of gravity and compresses the disc, thus compared with the non-magnetic case, in general the disc half-thikness, $\Delta\theta$, significantly is reduced. On the other hand, two parameters, appearing due to action of magnetic field and reaction of the flow, affect the disc thickness. The first one , $\beta_0$, showing the magnetic field strength at the equatorial plane, decreases $\Delta\theta$. The other one, $\eta_0$ is the magnetic resistivity parameter and when it increases, $\Delta\theta$ increases, too.
\end{abstract}


\section{INTRODUCTION}
Accretion onto black holes has been known as a powerful source of energy in the universe. Accreting gas with sufficiently high angular momentum tends to form a disclike structure around the central object. In accreting processes, viscosity causes angular momentum transport outward  and also it releases gravitational energy. According to the standard accretion disc model, the released energy is converted into radiation and escapes from the disc in the same place of its generation. The modern standard theory was formulated in Shakura (1972), Novikov \& Thorne (1973) and Shakura \& Sunyaev (1973). It provided remarkably successful contributions to understanding quasars, X-ray binaries and active galactic nuclei. One of the basic assumptions of this model is that the vertical thickness of the disc $H$ is much smaller than the corresponding radius $r$ in cylindrical coordinates ($H\ll  r$). Although the disc is geometrically thin, it is optically thick due to absorption and radiation completely moves outward in the vertical direction.

On the other hand, the energy released through viscosity may be trapped within accreting gas and then transported (advected) in the radial direction toward the central object or stored in the flow as entropy  (Narayan \& Yi 1995).  In this case the gas tends to have higher temperature which leads to a vertical thickening of the disc (H $\sim$ R).
 So in this situation, the accretion flow is called a radiatively inefficient accretion flow (RIAF) and depending on the mass accretion rate and optical depth of flow is divided into two types namely slim disc and ADAF (advection dominated accretion flow). The slim accretion disc model is introduced by Abramowicz et al. (1988) where flow is optically thick because of having large mass accretion rate. In the other type, ADAF, the accreting flow becomes optically thin in the limit of low mass accretion rate.  

For both branchs of hot accretion flows the solutions proposed by Narayan \& Yi (1994) is applicable. Using self-similar technique they assumed all variables to have power-law dependence on $r$ and then started to integrate the flow equations in the vertical direction. But vertical integration is valid for thin disc approximation only. Thus Narayan \& Yi (1995) retried to obtain a new solution in spherical coordinates to approach a more exact model. At that time the flow was considered completely thick and occupied the whole region between two poles. In both solutions the thickness value of the flow isn't exactly clear. 

It is generally believed that magnetic field has an fundamentally important role in the physics of accretion discs. For example MRI instability is known as a sure generator of turbulence in a Keplerian disc where angular momentum decreases outwards (Balbus \& Howly 1991). Indeed the crucial role of magnetic field in a hot flow is expected because of the high temperature of accreting gas in ADAFs ($10^9-10^{12}$ K). In this case the flow is ionized and may be influenced strongly in the presence of magnetic field.  For the first time Lynden-Bell (1969) considered the role of magnetic field in the context of active galactic nuclei and found how it might be responsible for angular momentum transport and the origin of anomalous disc viscosity. Bisnovatyi-Kogan and Blinnikov (1976) demonstrated that explanation of hard X-ray and gamma radiation from
Cyg X-1 would need presence of a magnetic field in the accretion disc. Some effort have been made to solve the magnetohydrodynamics (MHD) equations of magnetized ADAFs analytically. Kaburaki (2000) has presented a set of analytical solutions for a fully advective accretion flow in a global magnetic field. Shadmehri (2004) has extended this analysis for a non-constant resistivity. Abbassi et al. 2009 and Ghanbari, Salehi \& Abbassi (2007) have presented a set of self-similar solutions for two-dimensional (2D) viscousÐresistive ADAFs in the presence of a dipolar magnetic field of the central object. They have shown that the presence of a magnetic field and its associated resistivity can considerably change the picture with regard to accretion flows.

The effects of ordered magnetic fields in the accretion disc theories are often studied in two classes. In one class, the magnetic field is global and both poloidal and toroidal components of the ordered field are considered seriously. In the other class, the presence in the disc of only a toroidal field.
The latest case is acceptable since the dominant motion in an accretion disc has differential rotation, so it causes the toroidal component of magnetic field to become the most important one. Toroidal field is created by the action of differential rotation on initially poloidal field lines connecting layers rotating at different rates (Papaloizou \& Terquem 1997). In this class the magnetic field is often assumed to have even polarity which means being the same in the both sides of the equatorial plane and its effect is usually seen in the total pressure (i.e. gas plus magnetic). So in this view the behavior of magnetic and gas pressure are assumed to be the same and both support the disc against the vertical component of gravity. Therefore if the total pressure is substituted in the $\alpha$-prescription of viscosity, an additional viscous extraction of angular momentum passing through the disc plane becomes possible (Kato et al.2008). The effect of toroidal magnetic field on the disc was studied by Fukue (1990), Akizuki \& Fukue (2006), Abbassi et al. (2008), (2010) and Khesali \& Faghei (2011).

The global magnetic field could have odd or even-symmetry about the equatorial plane. Lovelace et al. (1986) and (1987) proposed a general theory for the axisymmetric flows around a black hole in a cylindrical $(r,\phi,z)$ coordinate system, and showed that in the presence of a magnetic field, the magnetic force can affect thickness of the accretion disc. Wang et al. (1990) considered a viscous resistive accretion disc in the presence of a global magnetic field. By using the thin disc approximation they have concluded that in odd-symmetry about the equatorial plane, $z=0$, ($B_z(r,z)=-B_z(r,-z),B_r(r,z)=B_r(r,-z), B_\phi(r,z)=B_\phi(r,-z)$ notice: here $B_\phi$ is even-symmetry ) the vertical component of magnetic force is opposed by gravity but in the case with even-symmetry the magnetic force is a compressive force like gravity. Nevertheless, if the magnetic field is purely toroidal with odd configuration, magnetic force will compress the disc (see Campbell \& Heptinstall 1998; Liffman \& Bardou 1999 for details).

The thickness of the advection-dominated disc isn't well defined. There is just a rough approximation used in the $\alpha$-prescription, (i.e. $H/r=c_s/v_k$, $c_s$ is the sound velocity at the disc equator and $v_K$ is the Keplerian velocity) in height integrated cylindrical coordinate solutions or even in spherical solutions presented by Narayan\& Yi (1995) (hereafter NY95).  Overcoming this problem, Gu \& Lu (2009) (hereafter GL09) introduced a somewhat different way to estimate the flow thickness. They did not give the value of $f=Q_{adv}/Q_{vis}$ ($Q_{adv}$ is the advective cooling rate per unit area and $Q_{vis}$,viscous heating rate per unit area) in advance, but instead considered accretion flows with free surfaces. The boundary condition is set to $p = 0$ which is usually adopted in the literature (e.g., Kato et al. 2008). Thus the thickness of the disc $\Delta\theta$ makes sense and they calculated $f$ to see how it relates to $\Delta\theta$. 

In this manuscript we aim to estimate the thickness of an advection-dominated disc in the presence of a purely toroidal magnetic field using GL09 method.

The outline of this paper is as follow: In Section 2 we present the basic magnetohydrodynamics equations, which include the induction equation with non-constant magnetic resistivity. Self-similar equations are investigated in section 3. In section 4, we will explain about the new lookout of advection parameter.  The result of the numerical solution and derive the disc thickness will present in section 5 and finally discussions and conclusions are in section 6.

\section{Basic Equations}

In this paper we consider a steady state ($\partial/\partial t=0$)  axisymmetric ($\partial/\partial \phi=0$) hot accretion flow. Spherical coordinates are used ($r, \theta, \phi$). The gravitational field is only emanating from a central point mass and we neglect self-gravity of the accreting flow. We also neglect the relativistic effects. The basic equations of the system are composed of continuity, momentum and induction equations. The equation of continuity is:
\begin{equation}
\frac{\partial\rho}{\partial t}+  \nabla \cdot (\rho\textbf{V})=0,
\end{equation}
the equation momentum conservation is:
\begin{equation}
\rho \frac{D\textbf{V}}{Dt}=-\nabla p - \rho\nabla\Phi+\textbf{F}^{\nu}+\frac{1}{c}(\textbf{J}\times \textbf{B}),
\end{equation}
Where $D/Dt=\partial/\partial t+\textbf{V}\cdot\nabla$.  And finally the Faraday's law of induction becomes
\begin{equation}
\frac{\partial\textbf{B}}{\partial t}=\nabla\times(\textbf{V}\times\textbf{B})-\nabla\times(\eta\nabla\times\textbf{B}).
\end{equation}
where $\rho$, $p$, $\textbf{v}$ and  $\textbf{B}$  are  the density of the gas, the pressure, the time-averaged flow's velocity
and the time-averaged magnetic field, respectively. These equations are supplemented by the Maxwell equations: $\nabla\times\textbf{B}=4\pi\textbf{J}/c,$ and by
$\nabla\cdot\textbf{B}=0.$ Here, $\eta$ is the magnetic diffusivity, $\textbf{F}^\nu=-\nabla\cdot\textbf{T}^\nu$
is the viscous force with $T_{jk}^\nu=-\rho\nu(\partial v_j/\partial x_k+\partial v_k/\partial x_j -(2/3)\delta_{jk}
\nabla\cdot\textbf{V})$ (in Cartesian coordinates), and $\nu$ is the kinematic viscosity.
We assume that only the $r\phi$-component of the viscous stress tensor, $ T_{r\phi}$ is important. 
 In the spherical coordinates, continuity and three components of momentum equation can be respectively written as:
\begin{equation}
\frac{1}{r^2}\frac{\partial}{\partial r}(r^2 \rho v_r)+\frac{1}{r \sin\theta}\frac{\partial}{\partial \theta}(\sin\theta \rho v_\theta)=0,
\end{equation}
\begin{displaymath}
    v_r\frac{\partial v_r}{\partial r}+\frac{v_\theta}{r}\frac{\partial v_r}{\partial\theta}-\frac{1}{r}(v_\theta^2+v_\phi^2)
    \end{displaymath}
    \begin{equation}
  \hspace{2.3cm}  =-\frac{GM}{r^2}-\frac{1}{\rho}\frac{\partial p}{\partial r}+\frac{1}{c\rho}(J_{\theta}B_\phi-J_{\phi}B_\theta)
\end{equation}
\begin{displaymath}
    \frac{v_r}{r}\frac{\partial(r v_\theta)}{\partial r}+\frac{v_\theta}{r}\frac{\partial v_\theta}{\partial\theta}
    -\frac{v_\phi^2}{r}\cot\theta
    \end{displaymath}
\begin{equation}
 \hspace{2.8cm}   =-\frac{1}{r\rho }\frac{\partial p}{\partial \theta}+\frac{1}{r c\rho }(J_\phi B_r-J_r B_\phi)
\end{equation}
\begin{displaymath}
\frac{v_r}{r}\frac{\partial(r v_\phi)}{\partial r}+\frac{v_\theta}{r \sin\theta}\frac{\partial }{\partial\theta}(\sin\theta v_\phi)
\end{displaymath}
\begin{equation}
\hspace{2.3cm} =\frac{1}{\rho r^3}\frac{\partial}{\partial r}(r^3 T_{r\phi})+\frac{1}{c\rho}(J_r B_\theta-J_\theta B_r),
\end{equation}
where $v_r , v_\theta,$ and $v_\phi$ are the three velocity components. Here the induction equation is considered. 

We suppose a toroidal magnetic field $\textbf{B}=B_\phi\hat{\phi}$ (that satisfies of $\nabla \cdot \textbf{B}=0$ with
 axisymmetric assumption), therefore the components of current density, $\textbf{J}$ become
\begin{displaymath}
J_r=\frac{c}{4\pi r}\frac{1}{ \sin\theta}\frac{\partial}{\partial\theta}(\sin\theta B_\phi),
\hspace{0.2cm} J_{\theta}=-\frac{c}{4\pi r}\frac{\partial}{\partial r}(r B_\phi),
\hspace{0.2cm}J_{\phi}=0,
\end{displaymath}

NY95 assumed $v_{\theta}=0$ which implies that no accretion material can evaporate as outflow from the discs.
Here following NY95 for simpilicity we adopt $v_{\theta}=0$ because at this stage we are interested in studying possible effects of B-field on the vertical structure of the discs although, in a more realistic picture, $v_\theta \ne 0$ should be taken into account (Xue \& Wang 2005). By substituting the current density relation, J,  the magnetic field $\textbf{B}=B_\phi\hat{\phi}$ and also $v_\theta=0$,  the continuity and momentum equations (4) - (7) will reduce to:
\begin{equation}
\frac{1}{r^2}\frac{\partial}{\partial r}(r^2 \rho v_r)=0,
\end{equation}
\begin{equation}
    v_r\frac{\partial v_r}{\partial r}-\frac{v_\phi^2}{r}=-\frac{GM}{r^2}-\frac{1}{\rho}\frac{\partial p}{\partial r}
    -\frac{1}{4\pi\rho }\frac{B_\phi}{r} \frac{\partial}{\partial r}(r B_\phi),
\end{equation}
\begin{equation}
    v_\phi^2 \cot\theta
  =\frac{1}{\rho }\frac{\partial p}{\partial \theta}+\frac{1}{4\pi\rho }[\frac{B_\phi}{\sin\theta}\frac{\partial}{\partial\theta}(\sin\theta B_\phi)],
\end{equation}
\begin{equation}
v_r\frac{\partial(r v_\phi)}{\partial r}
=\frac{1}{\rho r^2}\frac{\partial}{\partial r}(r^3 T_{r\phi}),
\end{equation}
The $r\phi$ component of viscous stress tensor is defined by $ T_{r\phi}=\rho\nu r \partial(v_\phi/r)/\partial r$. For the viscosity,$\nu$ , we will use $\alpha$-prescription so $\nu=\alpha c_s^2 r/v_k$, where $\alpha$ is the constant viscosity parameter, $c_s$ is the sound speed defined as $c_s^2=p/\rho,$ and $v_k^2=GM/r$ is the Keplerian velocity. 

Induction equation has  three components; only its azimuthal component is remained since toroidal magnetic field configuration is assumed.  Since we assume the steady flows then
  $\partial B_\phi/\partial t=0,$  so we have
\begin{equation}
\frac{1}{r}\frac{\partial}{\partial r}[\eta\frac{\partial(r B_\phi)}{\partial r}-r v_r B_\phi
]+\frac{1}{r^2}\frac{\partial}{\partial\theta}[\frac{\eta}{\sin\theta}\frac{\partial}{\partial\theta}(\sin\theta B_\phi)]=0,
\end{equation}
It is clear that the above equations are nonlinear and we aren't able to solve them analytically. Therefore, it is useful to have a simple means to investigate the properties of solutions.

 Self-similar method have been very useful in astrophysics and widely adopted in the astrophysical literature, since similarity assumption reduces the complexity
of the partial differential equations. This technique was applied by Narayan \& Yi (1994) in order to solve the system of height-averaged equations of a hot accreting flow. Then they investigated numerically the range of validity of their self-similar solutions. Their result show over a range of intermediate radii the numerical solution is close to the self-similar form, e.g. in a typical case this range is  $10 r_{in}<r<10^{-2} r_{out}$ , where $r_{in}$ is the inner edge and $r_{out}$  is the outer edge of the disc.

We will present self-similar solutions of these equations in the next section.

\section{Self-Similar Solutions}
The main equations are a set of coupled differential equations and thus they require to solve numerically. However, there is a powerful technique to give an approximate solution.
This powerful technique is self-similar method which is a dimensional analysis and scaling law which is a common tool in astrophysical fluid mechanics.
 Similar to NY95, we assume self-similarity in the radial direction so all types of velocities are proportional to
 $r^{-1/2}$ and for density $\rho \propto r^{-3/2}$ therefore gas and magnetic pressure must be $(p,B_\phi^2) \propto r^{-5/2}$ .
 If we adopt the above self-similar scaling, in fact, the radial dependencies of all physical quantities are canceled out and remain a set of
 equations which all quantities just are a function of $\theta$.
If we put these self-similar relations in the continuity equation, no new results will achieve, but other equations become:
\begin{equation}
	 v_\phi^2=v_k^2-\frac{1}{2}v_r^2-\frac{5}{2}\ c_s^2 -\frac{1}{4} c_A^2
\end{equation}
\begin{equation}
	 v_\phi^2 \cot\theta=\frac{1}{\rho}\frac{\partial
 (\rho c_s^2)}{\partial \theta}+c_A^2 \cot\theta+\frac{1}{2\rho}\frac{\partial (\rho c_A^2)}{\partial\theta}
\end{equation}
\begin{equation}
 v_r=-\frac{3}{2} \frac{\alpha c_s^2}{v_k}
\end{equation}

In the above three equations, the gas pressure was replaced by $4\pi\rho c_s^2$ , and the square Alfv\' en velocity, $c_A^2$, was used instead of $B_\phi^2/4\pi\rho $. Now, in order to appear terms including, $B_\phi^2$ and $B_\phi dB_\phi(=dB_\phi^2/2)$  in eqn. (12),  at first we multiply it to $B_\phi$, and then use self-similar relations, it becomes  
\begin{displaymath}
\frac{3}{4}(\frac{1}{4}\eta+r v_r )c_A^2+\frac{\partial\eta}{\partial\theta}(v_\phi^2 \cot\theta
-\frac{1}{\rho }\frac{\partial(\rho c_s^2)}{\partial \theta})
\end{displaymath}
\begin{displaymath}
 \hspace{1cm}+\eta[\frac{\partial v_\phi^2}{\partial\theta}\cot\theta-\frac{v_\phi^2}{\sin^2 \theta}-\frac{\partial}
 {\partial\theta}(\frac{1}{\rho }\frac{\partial(\rho c_s^2)}{\partial \theta})]
 \end{displaymath}
 \begin{equation}
\hspace{1.5cm}-\frac{\eta}{2}(v_\phi^2 \cot\theta-\frac{1}{\rho }\frac{\partial(\rho c_s^2)}{\partial \theta})
(\frac{1}{c_A^2}\frac{\partial c_A^2}{\partial\theta}
-\frac{1}{\rho}\frac{\partial\rho}{\partial\theta})=0,
\end{equation}

There are six unknown quantities: $\rho $, $ v_r , v_{\phi}, c _s^2 ,c_A^2$ and $\eta $ in four equations (13)-(16), so
we need two extra equations. One of them is a relation between pressure and density, and the other one is a relation for the resistivity. We assume a polytropic relation, $p=k\rho^\gamma$, in the vertical direction, (or equvalently meridional direction), where $\gamma$ is the ratio of specific heats; This is often adopted in the vertically integrated models of geometrically slim discs (e.g., Kato et al. 2008). We admit that the polytropic assumption is a simple way to close the system, and then enables us to calculate the dynamical quantities. So we can obtain:
\begin{equation}
\frac{1}{\rho}\frac{\partial\rho}{\partial\theta}=\frac{1}{(\gamma -1)c_s^2}\frac{\partial c_s^2}{\partial\theta}
\end{equation}

As mentioned above in order to complete the problem we need to adopt a physical assumption for magnetic diffusivity. We assume that the magnetic diffusivity is due to turbulence in the accretion flow and it is reasonable to express this parameter in analogy to the $\alpha$-prescription of Shakura and Sunyaeve (1973) for the turbulent viscosity, as follows (Bisnovatyi-Kogan \& Ruzmaikin 1976):
 \begin{equation}
 \eta=\eta_0 r \frac{c_s^2}{v_k},
 \end{equation}

Now, the system is completed, and it can be numerically solved with proper boundary conditions. We assume the structure of the disc is symmetric about the equatorial plane, and thus we have:
\begin{displaymath}
at \hspace{0.25cm}  \theta=\frac{\pi}{2}: \hspace{1.0cm} \frac{\partial c_s^2}{\partial\theta}=0, \hspace{0.1cm}\frac{\partial c_A^2}{\partial\theta}=0,
\end{displaymath}

We set $\rho(\theta=\pi/2)=1$ in order to obtain a unique solution by imposing a characteristic scale density at $\theta=\pi/2$. We need adopt proper values for $c_s^2$ and $c_A^2$ in equatorial plane. We expect the disc temperature is maximum in the equator of the disc, therefore $c_s^2(\theta=\pi/2)=c_{s0}^2$ must be maximum there and it will decreases toward the disc surface.  By integrating $\partial c_s^2/\partial\theta$ respect to $\theta$,  $c_s^2$ will decrease to reach zero in an angle $\theta_s$ for a given $c_{s0}^2$.  We have different $\theta_s$ for different $c_{s0}^2$. So as we expect hotter discs,  bigger values of $c_{s0}^2$ are thick vertically.  We need to fix $c_A^2$ at our inner boundary, $\theta=\pi/2$.  Now we may use the familiar relation between gas pressure and magnetic pressure ,that is 
 \begin{equation}
 \beta=2\frac{p_m}{p_g}=2\frac{B_\phi^2/8\pi}{p}=\frac{c_A^2}{c_s^2}
\end{equation}

We must emphasize that in this study $\beta$ is a function of $\theta$ since  $c_A^2$ and  $c_s^2$ are a function of $\theta$ while usually it was adopt a constant with respect to $\theta$
(Akizuki \& Fukue 2006, Abbassi et al. 2008). In the appendix, we have shown that $B_\phi^2$ is minimum at $\theta=\pi/2$. Since $c_s^2$, $B_\phi^2$ and automatically $c_A^2$ have contrary behavior respect to $\theta$,  $c_A^2$ increases from the equator toward the surface. Using the definition of $\beta$ we will be able to choose  reasonable boundary conditions
for  $c_A^2$ and  $c_s^2$.  If we consider equation (13) in $\theta = \pi/2$  using
the definition of $\beta$, (19) and eqn. (15) we have:
 \begin{equation}
	 v_{\phi0}^2=v_k^2-\frac{9\alpha^2}{8v_k^2}c_{s0}^4-(\frac{5}{2}+\frac{\beta_0}{4}) c_{s0}^2
\end{equation}
where $\beta_0=\beta(\theta=\pi/2)$ and zero index in the other quantities implies the value of them in the equator plane. It can be deduced admissible maximum value of $v_{\phi0}^2$ is $v_k^2$.
On the other hand, because of $v_{\phi0}^2$ must be positive, the right-hand-side of eqn (20) must also be positive. Hence we can determine an acceptable interval value of $c_{s0}^2$ per given $\beta_0$,  e.g. $0<c_{s0}^2<0.38v_K^2$ for a case $\alpha=0.1, \beta_0=0.3$, therefore  $c_{A0}^2=0.3c_{s0}^2$ is determined, but as we see the initial values of the both velocities are still arbitrary with just an upper limitation.
Now, with these proper boundary conditions, we are able to solve the main equations to conclude vertical behaviors
of the velocities and then we can determine how the magnetic field affects the disc thickness.

\section{Advection Parameter}

In previous section without applying of energy equation we determined velocities and pressures. In order to complete the equation it needs to have an energy equation. In this section we will focus energy transport equation. In principle, the general energy equation should be solved, and then advection parameter is obtained as a variable, as done by Manmoto et al. (1997) for ADAFs and by Abramowicz et al. (1988) and Watarai et al. (2000) for slim discs. Following Narayan \& Yi 1994 we adopt an advection form for energy equation ($Q_{adv}=Q_{+}-Q_{-}=fQ_{+}$) where $Q_{+},Q_{-}$ and $Q_{adv}$ are the heating rate per unit area, the cooling rate per unit area, and the advecting cooling rate per unit area, respectively. Here we introduce the advecting heating rate per unit volume as:
\begin{equation}
q_{adv}=\rho T \frac{Ds}{Dt}=\rho\frac{De}{Dt}-\frac{p}{\rho}\frac{D\rho}{Dt}
\end{equation}
Because we adopt a steady state, axisymmetric flows with $v_\theta=0$, we have $D/Dt=\textbf{V}\cdot\nabla=v_r\partial/\partial r$.
So $q_{adv}$ becomes:
\begin{equation}
q_{adv}=\frac{\rho v_r}{\gamma -1}\frac{\partial c_s^2}{\partial r}-c_s^2 v_r\frac{\partial\rho}{\partial r}
\end{equation}
 
On the other hand, we know $q_{adv}=q_+ - q_-$, where $q_+=q_{vis}+q_B$ is the dissipation rate per unit value. $q_{vis}=\eta r^2(d\Omega/dr)^2$, $q_B=J^2/\sigma$ are  generated energy due to the viscosity and magnetic resistivity, respectively, where $\sigma$ is conductivity of plasma; Here instead of it we use diffusivity, $\eta=c^2/4\pi\sigma$.  Further, in the self-similar formalism, they are simplified as:
  \begin{equation}
 q_{vis}=\frac{9\alpha}{4}\frac{pv_\phi^2}{rv_k},
 \end{equation}
 \begin{equation}
 q_B=\frac{\eta}{4\pi}|\nabla \times \textbf{B}|^2=\frac{\eta}{4\pi r^2}[(\frac{1}{\sin\theta}\frac{\partial}{\partial\theta}
 (\sin\theta B_\phi))^2+\frac{B_\phi^2}{16}]
 \end{equation}
also $q_-  = q_{rad}$ implies the energy lose through radiative cooling and we can write
 \begin{equation}
 q_{adv}=q_+-q_-=f' q_+
 \end{equation}
where $f'$ is the advection parameter which that shows what fraction of generated energy has remained in the disc at the definite polar angle. 
 \begin{equation}
 q_{adv}=-\frac{5-3\gamma}{2(\gamma-1)}\frac{pv_r}{r},
 \end{equation}
 
 The advection parameter $f$ shows whole energy trapping the entire disc thickness, it is expressed with $f=Q_{adv}/(Q_{vis}+Q_{B})$ where can be achieved by vertical integration over $q_{vis}, q_{adv}$ and $q_{B}$
  \begin{equation}
 Q_{adv}=\int_{\theta_s}^{\pi-\theta_s}q_{adv} r \sin\theta d\theta,
 \end{equation}
  \begin{equation}
 Q_{vis}=\int_{\theta_s}^{\pi-\theta_s}q_{vis} r \sin\theta d\theta,
 \end{equation}
\begin{equation}
 Q_B=\int_{\theta_s}^{\pi-\theta_s}q_B r \sin\theta d\theta,
 \end{equation}

In our system, we have seven physical quantities varying with polar angle: $v_r, v_{\phi}, p, \rho, c_s$, $B_{\phi}$ and $\eta$. For solving this equation to have these quantities,
we need seven equations , (13) - (19). Using proper boundary conditions which they are introduced in the previous section and with integrations respect to $\theta$, we will obtain numerically the vertical distribution of the above seven quantities.
 \section{Results}

 \subsection{Vertical Structure}
 \subsubsection{Notification about the Nonmagnetic Solutions}
  
In this section we will study the variation of dynamical quantities with the polar angle $\theta$ for a reasonable value of the squared sound velocity at the equatorial plane, $c_{s0}^2$ (or equivalently the gas temperature). But before that, we are interested in reviewing some details of the non-magnetic case. According to the original work of GL09 one boundary condition is required for solving equations, which was set to be $c_s=0$ at the surface of disc, $\theta=\theta_s$. Then we are able to find a solution for almost all given disc's half-opening angle, $\Delta\theta[=\pi/2-\theta_s]$ , but some of them which approach to nearly spherical configuration, aren't acceptable because of the limitation on advection parameter, $f$ which must be less than unit. Let us to notice the equation (10) of that paper and deduce $v^2_\phi$ from it and neglecting radial velocity comparing with other velocities we can conclude an approximate range for the value of the squared rotational velocity:
\begin{displaymath}
v^2_\phi\approx v^2_k-\frac{5}{2}c^2_s
\end{displaymath}  
from this equation we can determine an upper limit for $c^2_s$  it is $c^2_{smax}\approx 0.4 v^2_k$ that corresponds to $v_\phi^2\approx 0$. So there is an upper limit for gas pressure according to $p=\rho c_s^2$ so the maximum pressure at $\theta=\pi/2$ becomes $p_{max}\approx 0.4\rho_0 v^2_k$.  So with an acceptable initial value of $c^2_{s0}$ among 0 and $0.4v^2_K$ , we can start integrating from the equatorial plane instead of the surface of the disc (that was used by GL09), then we will conclude that maximum disc half-thickness belongs to a maximum value of $c^2_{s0}$. In this paper we follow this way and as we will see in the next subsection that starting integration from the other boundary is not applicable in this work at all.
 
\subsubsection{Solutions for the Magnetic Case}

Using the main equations and their boundary conditions that were introduced in last sections, equations (13)-(19), we can numerically derive the $\theta$-direction of distribution of physical quantities for a given radius. In our calculation we set $\gamma=3/2, \alpha=0.1$ , $\eta_0=0.1$ and $c_{s0}^2=0.1v_k^2$ and the behavior of the solution are investigated for different values of $\beta_0(=2p_m/p_g)$ (zero index means at $\theta=\pi/2$). For comparison, we have presented the nonmagnetic quantities with the black lines in figures 1,2 and 3.  
As Fig.1 to 3 represent the thickness of the disc decreases by increasing $\beta_0$. They show that when the magnetic field strength increases by adding $\beta_0$, the half-thickness of the discs will decrease considerably when we use identical initial condition (the same sound speed at $\theta=\pi/2$  consequently the same temperature in mid plane). So we conclude that the non-magnetized disc has the maximum thickness. Top panels of Figure 1 shows the variations of  $c_s$ and $c_A$ for some values of $\beta_0$. Obviously, they have contrary behavior respect to $\theta$ as we mentioned it in the previous section. The top left panel of Fig.1 displays the sound velocity does not has significantly change near the mid-plane but in the edge of the disc it decreases with increasing $\beta_0$ until attain zero in non-magnetic flow. From the top right panel of Fig.1 we see
 the Alfv\' en velocity  $c_A(\theta)$ is minimum at $\theta=\pi/2$ and is increasing slowly except near the surface. As this plot clearly
shows,  $\beta_0$ effects mainly near the edge. 

In this case the discs surface layers are non turbulent, and thus highly conducting (or non diffusive) because the
MRI is suppressed high in this situation (Lovelace et al. 2009). The physics of boundary layer of the discs and its corresponding physical phenomena is not the aim of this manuscript and it should be checked in our future investigation.
The bottom left panel shows the variation of the radial velocity, $v_r(\theta)$ with respect to the polar angle $\theta$ for the same given parameters. As we expect for ADAFs the radial velocity is sub-Keplerian, and absolute values of the radial velocity decreases with the vertical thickness of the discs and
reach to zero at the surface of the discs. When the magnetic parameter, $\beta_0$, becomes larger, the absolute value of $v_r$ decreases in a
given polar angle $\theta$ and magnitude of this reduction is more significant near
the edge of the discs. It reprint that that the maximum radial velocity is at equatorial region and toward the surface it tends to become zero.

The rotational velocity is shown in the bottom right panel of Fig1. It clearly shows can find $v_\phi$ is nearly independent of $\theta$ and even $\beta_0$ in the middle region of disc but near the edge it changes rapidly.  For non-magnetic case we can see the rotational velocity behaves such as a monotonic function of $\theta$ it is monotonically increasing from $\theta=\pi/2$ to $\theta=\pi/2\pm\Delta\theta$ . But when the magnetic field plays an important role changes differently. The magnetic field divide the disc into the two distinct regions, one region is around the mid-plane where the rotational velocity changes slowly. The other one is near the surface $v_\phi$ is maximum at first and then decreases very fast and finally reaches to zero on the surface. On the other hand, with decreasing $\beta_0$ the surface layer $\Delta\theta_s$, becomes thinner so for the limit of $\beta_0=0$ it tends to $\Delta\theta_s=0$ . But it happens on the surface boundary where the conditions are so different from the inner region, therefore we aren't able to have an accurate solutions there. However, a mass point  rotates slower in a magnetized disc because it has experienced an extra magnetic force. As we expected the radial and rotation velocities are both sub-Keplerian. Fig2. shows the gas and magnetic pressures scaled with a fiducial pressure $p_0=\rho_0 v^2_k$. The gas pressure as we expected behaves as the same as the radial velocity. it peaks at  $\theta=\pi/2$ and in a weak magnetic field is almost constant but magnetic pressure, i.e. $p_m$ in the right of Fig.2, is minimum at the equatorial plane and becomes larger toward the surface. It shifts up when the magnetic parameter $\beta_0$ increases. 

  \input{epsf} 
\epsfxsize=3.5in \epsfysize=2.5in
 \begin{figure}
\centerline{\epsffile{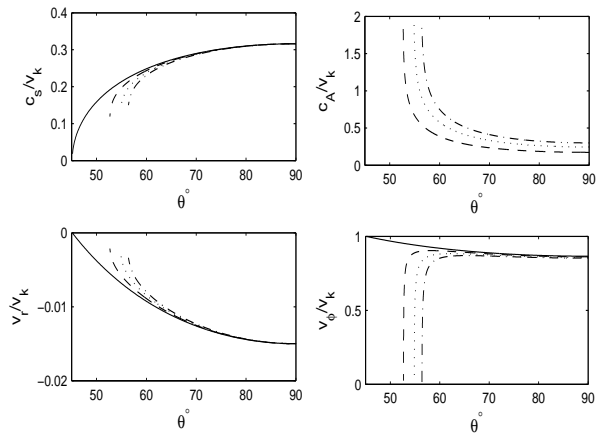}}
\caption{Self-similar solutions corresponding to $\gamma=3/2$, $\alpha=0.1$, $\eta_0=0.1$ and several values of $\beta_0$. The black line  shows solution at the non-magnetic situation and also dashed, dotted and dot-dashed
 line refer to $\beta_0=0.3, 0.6, 0.9$ and the solid line shows the corresponding quantity in the absence of magnetic field. In this figure, $v_k$ is the Keplerian velocity.}  
 \label{figure1 }
\end{figure}

\input{epsf} 
\epsfxsize=3.5in \epsfysize=1.1in
 \begin{figure}
\centerline{\epsffile{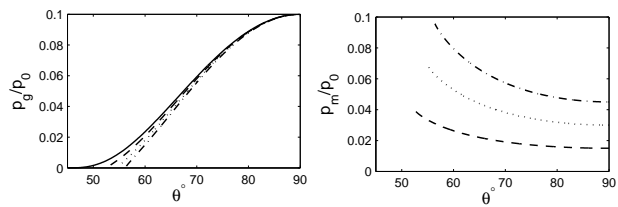}}
\caption{Self-similar solutions corresponding to $\gamma=3/2$, $\alpha=0.1$, $\eta_0=0.1$ and several values of $\eta_0$. The dashed, dotted and dot-dashed line refer to $\beta_0=0.3, 0.6, 0.9$, and the solid line shows the corresponding quantity in the absence of magnetic field. In this figure, the fiducial pressure $p_0$ is determined by $p_0=\rho_0 v^2_k$}.
 \label{figure2 }
\end{figure}

 \input{epsf} 
\epsfxsize=3.5in \epsfysize=1.2in
 \begin{figure}
\centerline{\epsffile{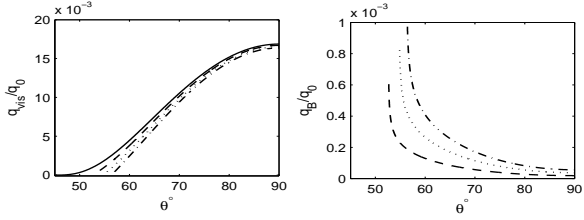}}
\caption{Variation of two forms of the dissipated energy ($q_{vis},q_B$) per unit volume of plasma, with $\gamma=3/2$, $\alpha=0.1$, $\eta_0=0.1$ and several values of $\beta_0$. The dashed, dotted and dot-dashed line refer to $\beta_0=0.3, 0.6, 0.9$ and the solid line shows dissipated energy by viscosity in the non-magnetic flow. In this figure, the fiducial dissipated energy is $q_0=p_0v_K^2/r$, where $p_0=\rho_0 v_K^2$.}
 \label{figure3 }
\end{figure}

Fig. 3 shows variations of the scaled viscose and magnetic energy dissipation in two separate panel for the same parameters ($\gamma=3/2$, $\alpha=0.1$, $\eta_0=0.1$). It clearly shows that when the toroidal magnetic field becomes stronger, the viscous dissipation decreases. But according to the right panel of this figure, the heating process by resistivity is grown up in a stronger magnetic field.  Moreover, maximum value of $q_B$ which is happend at the surface of disc, is almost one order of magnetude less than viscosity dissipation at the midplane, means $q_{B_{max}}\approx0.1q_{vis_{max}}$. Although, the ressistive dissipation can not have much effect on the advection parameter,  it can make $f'$ significantly change at the outer parts of the disc. 
\subsubsection{The Role of the Resistivity}

One of the prominent input parameters in our system is the magnetic diffusivity , $\eta_0$,  that its possible effects are explored in Figs.4-6. We assume that $\gamma=3/2$, $\alpha=0.1$ and $\beta_0=0.5$. The solid, dash and dot lines correspond to $\eta_0 =0.05, 0.1 , 0.2$ respectively. Fig. 4 displays sound speed $c_s$ (top, left) , rotational velocity $v_\phi$ ( bottom, right) which they are normalized by Keplerian velocity. They are constant near the mid-plane but begin to change toward the surface. Radial velocity $v_r$ is shown in the bottom left panel of fig4. at a fixed equator temperature but for different values of  $\eta_0$; it has minimum value at $\theta=\pi/2$ and  $\eta_0$ affects on the redial velocity mainly near the edge of the disc.  As it clearly shows for higher value of resistivity parameter total velocity increases in a disc.  On the contrary, in a given $\theta$ an increase in $\eta_0$ leads to  a decrease in Alfv\' en velocity $c_A$ (top, right).

Fig. 5 gives the effect of resistivity parameter on the gas and magnetic pressure. The left panel shows that for higher values of $\eta_0$, the gas pressure rises in the outer regions but the magnetic pressure that is shown in the right of Fig. 5 diminishes. Finally in Fig. 6 we can see how the magnetic diffusivity affects on two heating energy sources, $q_{vis}$ (is shown at left) and $q_B$ (at right). The magnetic resistive affects on $q_B$ explicitly, but it affect on $q_{vis}$  implicitly throughout the dynamical quantities.
 According to this figure, viscosity dissipation is an ascending function of $\eta_0$ in the outer regions. In spite of,  $q_B$ is proportional to  $\eta_0$ , it falls off as $\eta_0$ becomes larger and for small values of $\eta_0$ it tends to be constant along $\theta$ direction. As we see Ohmic heating is so smaller  than viscous heating so we can ignore the role of magnetic field in energy equation. 
  \input{epsf} 
\epsfxsize=3.5in \epsfysize=2.2in
 \begin{figure}
\centerline{\epsffile{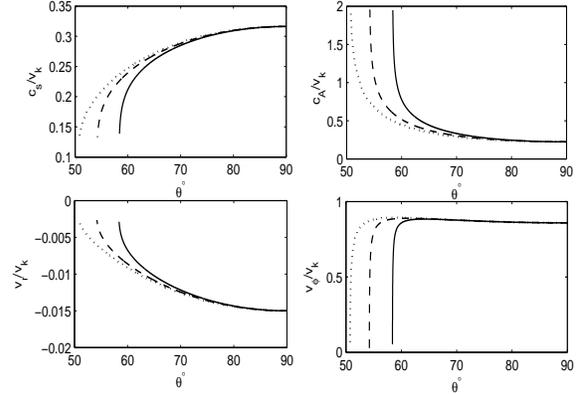}}
\caption{Self-similar solutions corresponding to $\gamma=3/2$, $\alpha=0.1$, $\beta_0=0.1$ and several values of $\eta_0$. The solid , dash and dot line refer to $\eta_0=0.05, 0.1, 0.2$ . $v_k$ is the Keplerian velocity .  }
 \label{figure4 }
\end{figure}
  
\input{epsf} 
\epsfxsize=3.6in \epsfysize=1.2in
 \begin{figure}
\centerline{\epsffile{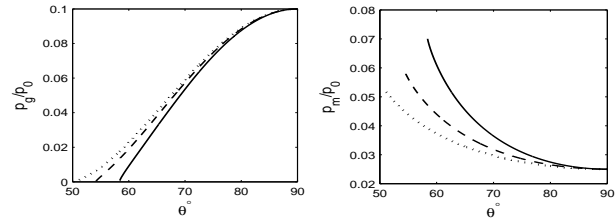}}
\caption{The profiles of the gas and magnetic pressure corresponding to $\gamma=3/2$, $\alpha=0.1$, $\beta_0=0.5$ and several values of $\eta_0$. The solid , dash and dot line refer to $\eta_0=0.05, 0.1, 0.2$. In this figure, the fiducial pressure $p_0$ is determined by $p_0=\rho_0 v^2_k$}.
 \label{figure5 }
\end{figure}
  \input{epsf} 
\epsfxsize=3.5in \epsfysize=1.2in
 \begin{figure}
\centerline{\epsffile{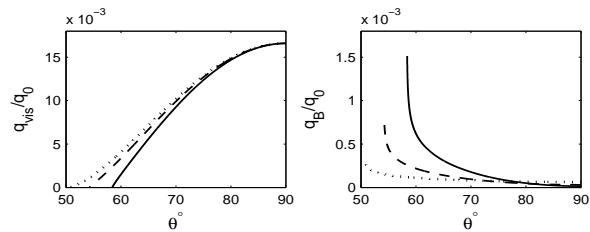}}
\caption{Variation of two forms of the dissipated energy ($q_{vis},q_B$) per unit volume of plasma, corresponding to $\gamma=3/2$, $\alpha=0.1$, $\beta_0=0.5$ and several values of $\eta_0$. The solid , dash and dot line refer to $\eta_0=0.05, 0.1, 0.2$. In this figure, the particular dissipated energy is $q_0=p_0v_K^2/r$, where $p_0=\rho_0 v_K^2$.} 
 \label{figure6 }
\end{figure}

\subsection{ The Thickness of Flow  }

In this section we will explore that how the physical input parameters of the system affect on the thickness of flow. 
At first, we will review the way that is usually used for $\alpha$ discs and then explain about the especial way applied by GL09  with more details  for the magnetized case.
\subsubsection{The Usual Way for Estimating of the Flow Thickness  }
 At first we consider the usual approximation for the thickness of flow that is based on $\alpha$-prescription, means $\nu=\alpha c_s H$ where $H=c_s/\Omega_K$ or  $H/r=c_s/v_K$ and H is the half-thickness of the disc. This is result of hydrostatic equilibrium; means the gravitational force and the pressure force are balanced with each other in the vertical direction. 
 \begin{equation}
 \frac{1}{ \rho}\frac{\partial p}{\partial z} +\frac{\partial\psi}{\partial z} =0
 \end{equation}
where $\psi$ is the gravitational potential in a cylindrical coordinate can be written as: $\psi=-GM/(r^2+z^2)^{1/2}$. Now using some approximations we can estimate the half-thickness of disc in this way: $\partial p/\partial z\approx - p_0/H, $  $\rho\approx \rho_0$ (zero index shows the value of quantity at the equatorial plane) $\partial\psi/\partial z \approx GMH/r^3=\Omega^2_K H$ and the sound velocity is $c_{s0}=(p_0/\rho_0)^{1/2}$. As we mentioned at subsection 5.1.1, the squared of sound velocity can not exceed more than $0.4 v^2_K$ so $H/r\leq\sqrt{0.4}=0.63$  in the non-magnetic flow. 

In the presence of a toroidal magnetic field, the magnetic force is added to the vertical component of motion equation:
 \begin{equation}
 \frac{\partial p}{\partial z} +\frac{B_\phi}{4\pi}\frac{\partial B_\phi}{\partial z}+ \rho\frac{\partial\psi}{\partial z} =0
 \end{equation}   

We can use the magnetic pressure $p_m=B^2_\phi/8\pi$, which is usually (e.g. Narayan, Yi 1995) assumed to be proportional to the gas pressure $p$, means $p_m=\beta p$ ,
so with previous assumptions, one can easily achieve $H/r=(1+\beta)^{1/2} c_s/v_K$. Therefore the half-thickness of the disc increases when we add the influence of magnetic field on the structure of accretion flow. We will show that this concolution is not valid indeed if we take into account the induction equation. In the following subsection we study several details which can affect on the thickness of disc in the presence of magnetic field. For simplification as we noted in the introduction in this study a toroidal configuration for magnetic field is assumed.

\subsubsection{ The Thickness of the Magnetized Flow}

As we noted in the introduction the aim of this manuscript is revisiting the vertical structure of hot accretion flow when magnetic field has an important role. In the other hand we expand the equations in spherical coordinate. So the study of vertical structure in magnetized case is a quite complicated and different from non-magnetized case since induction equation should be take into account.

As we mentioned in subsection 5.1.1, the solution shows that  the sound velocity decreases from the mid-plane toward the surface and it becomes zero at  the surface of non-magnetized disc. But in the presence of the magnetic field, at first ,we need to consider eqn. (13):
\begin{equation}
 v_\phi^2=v_k^2-\frac{1}{2}v_r^2-\frac{5}{2}\ c_s^2 -\frac{1}{4} c_A^2
\end{equation}
so the rotational velocity depends on not only $c_s$ but also $c_A$. Here, Alfv\' en velocity has crucial role in the determination of the disc thickness. As we saw in previous section in figures 1 and 4 , the sound and Alfv\' en velocities have different behaviors; therefore $v _\phi$ can have non-monotonic behavior from the equatorial plane toward edge the flow. At first, it increases because of decreasing $c^2_s$ but  $c^2_A$ is still too small to affect considerably on $v_\phi^2$. In somewhere between the mid-plane and the surface, $c^2_A$ becomes large enough, comparable with $c^2_s$, and makes the rotational velocity start to reduce until being zero at the disc surface. Thus, we must consider influences associated with Alfv\' en velocity's behavior. 

Although there are strict limitations for discussing in numerical solutions, we can consider a special point help us to determine the general behavior of the solution. The symmetry assumption about the equatorial plane hints this point that is $\theta=\pi/2$. There are two possible symmetry configurations for the toroidal magnetic field respect to equatorial plane, even and odd symmetry. Nevertheless, both of them lead to the same result, because $B^2_\phi$ is important here and it is even function and also minimum for both symmetries.  

The behavior of the magnetic field inside the disc depends on $\partial^2 B_\phi^2/\partial\theta^2=\Delta B^2_\phi$  at the equatorial plane. From the induction equation (see appendix), we can obtain:
\begin{equation}
\frac{\partial^2 B_\phi^2}{\partial\theta^2}\bigg|_{90^\circ}=(\frac{13}{8}+\frac{9\alpha}{4\eta_0}) B_{\phi0}^2 >0
 \end{equation}
 
 The above relation indicates that $B_\phi^2$  is minimum at the mid-plane. On the other hand, if  $\Delta B^2_\phi$ becomes larger,  $B^2_\phi$ will increase more rapidly and make the disc thiner. According to the last relation, $\Delta B^2_\phi$ is directly proportional to $B_{\phi0}^2=8\pi\beta_0c^2_{s0} $ and depends on directly $\alpha$ but inversely $\eta_0$ . 

Having inverse relationship between the disc thickness and $\Delta B^2_\phi$, an increasing in $\beta_0$ or $\alpha$ leads an decreasing in disc thickness but it increases by increasing $\eta_0$.   

We can see the effects of $\beta_0$  and  $\eta_0$ in Fig.7 , 8 . As we expect , Fig. 7 shows that by increasing the $\beta_0$ parameter,  the disc thickness decreases and this effect is stronger for the high temperature of the flow. However, we can say magnetic force in the vertical direction compresses the disc.  Liffman \& Bardou (1999) and Campbell \& Heptinstall (1998) also noted compression of disc in height direction by effect of toroidal magnetic field. From Fig.8, it is seen that the magnetic resistiviy has an direct effect on the thickness of the flow, when it is increased , the disc thickness also increased. 
\input{epsf}
\epsfxsize=2.5in \epsfysize=1.8in
\begin{figure}
\centerline{\epsffile{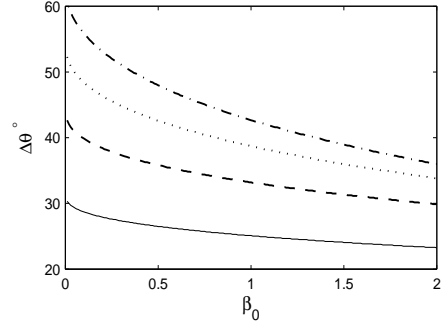}}
 \caption{The disc's half-opening angle, $\Delta\theta$, as a function of the strength magnetic field parameter at the equatorial plane, $\beta_0$, for $\gamma=3/2$, $\alpha=0.1$, $\eta_0=0.1$ and various values of the equatorial square sound velocity, the solid, dashed, dotted and dot-dashed lines represent  $c^2_{s0}/v^2_K=0.05, 0.10, 0.15, 0.20 $.  }
 \label{figure7 }
\end{figure}
\input{epsf}
\epsfxsize=2.5in \epsfysize=1.8in
\begin{figure}
\centerline{\epsffile{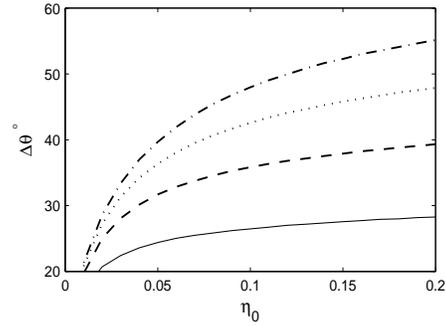}}
 \caption{The disc's half-opening angle, $\Delta\theta$, as a function of the resistivity parameter , $\eta_0$, for $\gamma=3/2$, $\alpha=0.1$, $\beta_0=0.5$ and various values of the equatorial square sound velocity, the solid, dashed, dotted and dot-dashed lines represent  $c^2_{s0}/v^2_K=0.05, 0.10, 0.15, 0.20 $.    }
 \label{figure8 }
\end{figure}
  
   \subsection{Advective Parameter }

As we mentioned before, in this work for determination of advective parameter at first we solve the system of differential equations and after finding dynamical quantities in the fluid we can specify $f$ according to equations (27), (28), (29) and this relation $f=Q_{adv}/(Q_{vis}+Q_B)$. 
It is seen that in existence of magnetic field that tends to compress the fluid, much more advecting energy can save in a disc with less thickness in comparing with non-magnetic flow. For example we can see (for$\gamma=4/3$) $f=0.1$ in a nonmagnetic flow with $\Delta\theta=0.4\pi=72^\circ$ while it happens in a thinner magnetized flow with $\Delta\theta=0.33\pi=59.4^\circ$ and $\beta_0=0.1$. So it helps somewhat that previous imagination about slim discs that they are no thin nor thick is retained its validity.

\input{epsf}
\epsfxsize=3.5in \epsfysize=1.5in
\begin{figure}
\centerline{\epsffile{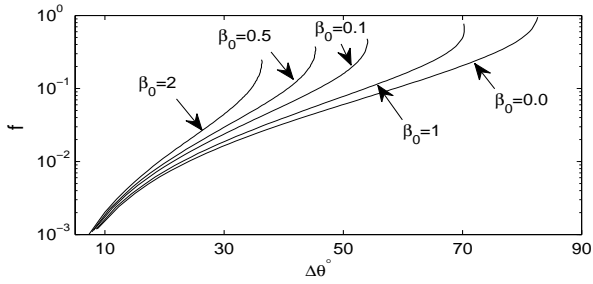}}
 \caption{Variation of the advection factor, $f$ with the disc's half-opening angle, $\Delta\theta$ for $\gamma=3/2$, $\alpha=0.1$, $\eta_0=0.1$ and various value of $\beta_0$.  }
 \label{figure9 }
\end{figure}

\input{epsf}
\epsfxsize=3.5in \epsfysize=1.5in
\begin{figure}
\centerline{\epsffile{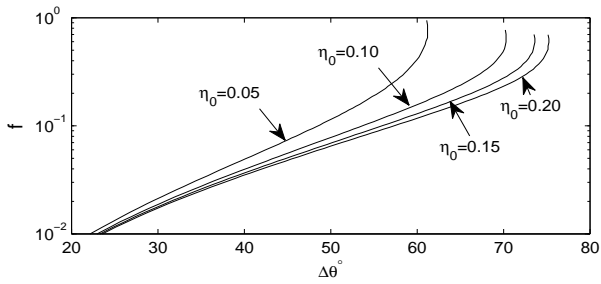}}
 \caption{Variation of the advection factor, $f$ with the disc's half-opening angle, $\Delta\theta$ for $\gamma=3/2$, $\alpha=0.1$, $\beta_0=0.1$ and various value of $\eta_0$  . }
 \label{figure10 }
\end{figure}

In figure 9, $f$ is plotted as function of $\beta_0$ for the fixed value of $c_{s0}$, it shows that much more energy can be advected in a stronger magnetic field that exists in a thinner disc; and in a hotter disc (that is, greater sound velocity) advection will be intensified.

The profile of energy advection factor, $f$, for various values of magnetic resistivity parameter $\eta_0$ is presented as a function of the disc's half-opening angle in Fig. 10.  It demonstrates that energy advection decreases inversely with $\eta_0$ in a fixed disc thickness.
 
\subsection{Bernoulli Parameter}

In stationary, inviscid flows with no energy sources or losses, the quantity (Abramowicz et. al. 2000)
\begin{equation}
Be_0=W+\frac{1}{2}V^2+\Phi
\end{equation}
is constant along each individual streamline, but, in general, is
different for different streamlines. This quantity is called the
Bernoulli constant. Here W is the specific enthalpy, V is the
velocity (all three components included) and $\Phi$ is the gravitational
potential.
\begin{equation}
Be_0=\frac{1}{2}(v^2_r+v^2_\phi)-\frac{GM}{r}+\frac{\gamma}{\gamma-1}\frac{p}{\rho}
\end{equation}

Obviously, a particular streamline may end up at infinity
only if $Be_0> 0$ along it. The existence of streamlines with $Be_0> 0$
is therefore a necessary condition for outflows in stationary
inviscid flows with no energy sources or losses, and $Be_0< 0$ for all
streamlines is a sufficient condition for the absence of outflows.
However, $Be_0> 0$ is not a sufficient condition for outflows.
In all viscous flows, $Be_0$
isn't constant along individual streamlines. For the self-similar solutions $Be_0$ is the function of $r^{-1}$ so it can not be a constant value at all. Hence, the so-called
`Bernoulli parameter', is introduced $\widetilde{Be}_0=Be_0/V^2_k$:

In the presence of a magnetic field, an extra term needs to be added to the Bernoulli function (Fukue 1990). Since the Bernoulli equation is based on energy conservation along each streamline, in the magnetic case, the total energy of the fluid is included of  the magnetic energy in addition to the other previous form of energies which are in eqn. (35). So the Bernoulli function of magnetized flow becomes: 
\begin{equation}
Be(r,\theta)=\frac{1}{2}(v^2_r+v^2_\phi)-\frac{GM}{r}+\frac{\gamma}{\gamma-1}\frac{p}{\rho}+\frac{B_\phi^2}{4\pi\rho}
\end{equation}
for the self-similar model it is simplified $Be(r,\theta)=Be(\theta)v^2_k$:
\begin{equation}
Be(\theta)=\frac{1}{2}[v^2_r(\theta)+v^2_\phi(\theta)]-1+\frac{\gamma}{\gamma-1}c^2_s(\theta)+c^2_A(\theta)
\end{equation}
\input{epsf}
\epsfxsize=2.75in \epsfysize=1.7in
\begin{figure}
\centerline{\epsffile{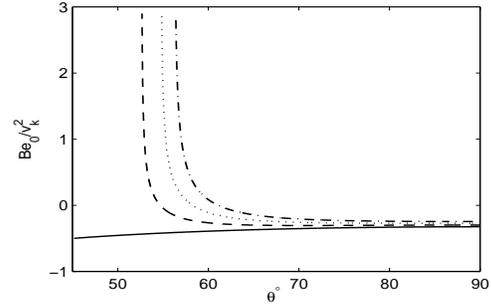}}
 \caption{Bernoulli parameter , with respect to $\theta$  for different value of $\beta_0$.  The black line  corresponds to $\beta_0=0$  dashed line $\beta_0=0.3$ , dotted line is for $\beta_0=0.6$ and dot-dashed line for $\beta_0=0.9$  ($c^2_{s0}=0.1v^2_k$ is the boundary condition at $\theta=\pi/2$ and $\gamma=3/2$). }
 \label{figure11 }
\end{figure}

As we see from Fig. 11, without magnetic field the Bernoulli function in ADAFs with low viscosity is negative, but in the presence of magnetic field it can achieve a positive value close to the surface. It means that GL09 solutions without magnetic field is not able to describe the existence of wind and outflow. 
  
\section{discussion and conclusion}

The vertical structure of a hot accretion flow is still an open problem. Hence in this paper following the work of GL09 we have considered a two-dimensional axi-symmetric advection-dominated accretion flow in spherical coordinates with a toroidal magnetic field. We have concentrated on studying possible effects of magnetic field and its corresponding resistivity on the radial and vertical accretion structure. With a self-similar solution along the radial direction and the proper boundary conditions using reflection symmetry in equatorial plane of the disc we have constructed the structure of the disc along the $\theta$ direction explicitly.

In this paper we used the induction equation for a resistive flow in order to complete the system of the basic equations of fluid dynamics. We assumed $\beta[=p_g/(p_g+p_m)]$ to be a function of $\theta$ while in the previous papers (Akizuki \& Fukue 2006, Abbassi et al. 2008, 2010) it was adopted as a constant. As a result we could find new solutions for the dynamical quantities of ADAFs in the presence of a toroidal magnetic field, $B_\phi$. The stationary solutions we found indicate that even a weak toroidal field at the mid-plane can grow significantly near the edge and cause a significant change in surface layers. Moreover, the main purpose of this paper is to investigate how the vertical thickness of the disc will change  in the presence of toroidal magnetic field, $B_\phi$. We considered an even configuration of $B_\phi$ and found that it has a squeezing effect on the disc structure, where the disc thickness is reduced compared to the non-magnetic case. In fact, the squeezing effect of $B_\phi$ counterbalances the thickening of the disc generated by advection (Shadmehri \& Khajenabi 2005). Our conclusions are opposite those of Wang, Sulkanen, and Lovelace (1990) (for the odd symmetry case) because the envisioned models are different in the two cases. Wang et al. assume a relatively thin disc $(H < r)$ whereas we consider a thick disc H $\sim$ R. The solution presented in this manuscript is in good agreement with that presented by  Mosallanezhad, Abbassi \& Beiranvand 2013 for which they used the same physical method and assumptions. Their solutions indicate that the outflow region, where the radial velocity becomes positive in a certain inclination angle $ \theta_{0} $, always exist. They have shown that a stronger toroidal magnetic field, lead a smaller inclination angle, which means a thinner disc.

The complex behavior of the flow depends on the input parameter of the problem and is explored in detail in this manuscript. However a complete analysis is needed to complete our model, including detailed analysis of the edge of the disc. However, we point out that the accretion outflow solutions are unstable near the outer edge and outside the accretion flow. Further and more detailed study should be done on the wind region solution and its interaction with Large-scale magnetic field. The present result lend strong support to the suggestion that magnetic field has an important role in the vertical structure of hot flows and magnetically channeled wind. Also we have noted that the diffusion properties of the magnetically dominated corona have never been investigated and it would be good if it is tested in future investigations.

The author is particularly grateful to Richard Lovelace. Mohsen Shadmehri and Fu-Guo Xi for their discussions and useful suggestions. We are also like to appreciate the referee for his/her thoughtful and constructive comments which clarify some points in the early version of the paper. This work was supported by Ferdowsi University of Mashhad under the grant 3/27130 (1392/03/16).

\appendix

\section{}
We seek a proper boundary for $c_A^2$ but before that, we need to
investigate behavior of $c_A^2$ from equatorial plane toward
disc surface, for this purpose we refer to the induction equation .
\begin{equation}
\frac{\partial B_{\phi}}{\partial t}=\frac{1}{r}\frac{\partial}{\partial r}[\eta\frac{\partial}{\partial r}
(r B_{\phi})-r v_r B_{\phi}]+\frac{1}{r^2}\frac{\partial}{\partial\theta}[\eta(B_{\phi}\cot\theta+\frac{\partial B_{\phi}}{\partial\theta})]
\end{equation}
It is supposed that $\partial B_{\phi}/\partial t=0$ and then by multiplying to $B_{\phi}/4\pi\rho$ we have:
(after multiplying it to $B_\phi$ and some simplification)
\begin{displaymath}
\frac{3}{8}(\frac{1}{2}-\frac{3\alpha}{\eta_0})c_s^2 B_\phi^2+\frac{\partial c_s^2}{\partial\theta}(B_\phi^2\cot\theta
+\frac{1}{2}\frac{\partial B_\phi^2}{\partial\theta})
\end{displaymath}
\begin{equation}
 +c_s^2[-\frac{B_\phi^2}{\sin\theta^2}+\frac{1}{2}\frac{\partial B_\phi^2}{\partial\theta}\cot
 \theta+\frac{1}{2}\frac{\partial^2 B_\phi^2}{\partial\theta^2}-\frac{1}{4B_\phi^2}(\frac{\partial B_\phi^2}{\partial\theta})^2]=0
\end{equation}
This relation in the equatorial plane ($\theta=\pi/2$) , (where $\partial/\partial\theta=0$) converts to
\begin{equation}
\frac{\partial^2 B_\phi^2}{\partial\theta^2}=\frac{13}{8}(1+\frac{18\alpha}{13\eta_0})B_\phi^2 >0
 \end{equation}
 So it is obvious $B_\phi^2$ must be minimum in the equator of disc.

\begin{thebibliography}{}
\bibitem[]{}Abbassi S., Ghanbari J., Najjar S., 2008, MNRAS, 388, 663
\bibitem[]{}Abbassi S., Ghanbari J., Ghasemnezhad M., 2010, MNRAS, 409, 1113
\bibitem[]{}Abramowicz, M. A., Czerny, B., Lasota, J. P., \& Szuszkiewicz, E. 1988, ApJ, 332, 646
\bibitem[]{}Abramowicz, M. A., Lasota, J.-P., \& Igumenshchev, I.V. 2000, MNRAS, 314, 775
\bibitem[]{}Akizuki, C.,  Fukue, J., 2006, PASJ, 58, 469
\bibitem[]{}Balbus, S. A., Hawley J. F., 1991, ApJ, 376, 214
\bibitem[]{}Bisnovatyi-Kogan. G.S., Ruzmaikin, A.A., 1976, Ap\& SS, 42,401 
\bibitem[]{}Bisnovatyi-Kogan. G. S., Blinnikov, S.I. 1976, Sov. Astron. Lett., 2, 191
\bibitem[]{}Campbell, C. G., Heptinstall, P., 1998, MNRAS, 299, 31
\bibitem[]{}Fukue, J., 1990, PASJ, 42, 793
\bibitem[]{}Ghanbari, J., Salehi, F., \& Abbassi, S., 2007, MNRAS, 381, 159
\bibitem[]{}Gu, W.-M., Xue, L., Liu, T., \& Lu, J.-F. 2009, PASJ,61, 1313
\bibitem[]{}Kaburaki, O. 2000, ApJ, 531, 210
\bibitem[]{}Kato, S., Fukue, J., \& Mineshige, S. 2008, Black-Hole Accretion disks - Towards a New Paradigm (Kyoto: Kyoto Univ. Press)
\bibitem[]{}Khesali A., Faghei K., 2009, MNRAS, 398, 1361
\bibitem[]{}Liffman,K., Bardou, A., 1999, MNRAS, 309, 443
\bibitem[]{}Lovelace, R. V. E., Mehanian, C., Mobarry, C. M., \& Sulkanen, M. E. 1986,ApJS, 62, 1
\bibitem[]{}Lovelace, R. V. E., Wang, J. C. L., \& Sulkanen, M. E. 1987,ApJ, 315, 504
\bibitem[]{}Lovelace, R. V. E., Romanova, M. M., \& Newman, W. I. 1994,ApJ, 437, 136
\bibitem[]{}Lovelace, R. V. E., Rothstein, D. M., \&  Bisnovatyi-Kogan, G. S. 2009, 701, 885
\bibitem[]{}Lynden-Bell, D., 1969, Nat, 223, 690
\bibitem[]{}Manmoto, T., Mineshige, S., \& Kusunose, M. 1997, ApJ, 489, 791
\bibitem[]{}Mosallanezhad A., Abbassi S., Beiranvand N., Submitted to MNRAS 
\bibitem[]{}Narayan, R., Yi, I., 1994, ApJ, 428, L13
\bibitem[]{}Narayan, R., Yi, I., 1995a, ApJ, 444, 231
\bibitem[]{}Narayan, R., Yi, I., 1995b, ApJ, 452, 710
\bibitem[]{}Novikov I.D., Thorne K.S. 1973, in Black Holes eds. C.DeWitt \& B.DeWitt (New York: Gordon \& Breach), p.345
\bibitem[]{}Papaloizou, J. C. B., Terquem, C. 1997, MNRAS, 287, 771
\bibitem[]{}Shadmehri, M., 2004, A\&A, 424, 379
\bibitem[]{}Shadmehri, M., Khajenabi, F. 2005, MNRAS, 361, 719
\bibitem[]{}Shakura N.I. 1972, Astron. Zh., 49, 921 (1973, Sov. Astron., 16, 756)
\bibitem[]{}Shakura, N.I., \& Sunyaev, R.A. 1973, A\& A, 24, 337
\bibitem[]{}Watarai, K.-Y., Fukue, J., Takeuchi, M., \& Mineshige, S. 2000, PASJ,52, 133
\bibitem[]{}Wang, J. C. L., Sulkanen, M. E., \& Lovelace, R. V. E. 1990,ApJ, 355, 38
\bibitem[]{}Xue, L., Wang, J. 2005, ApJ, 623, 372
\end{thebibliography}
\end{document}